# On Blue Straggler Formation by Direct Collisions of Main-Sequence Stars

James C. Lombardi, Jr.[1,2], Frederic A. Rasio[3,4] and Stuart L. Shapiro[1,2,5]

## ABSTRACT

We report the results of new SPH calculations of parabolic collisions between main-sequence (MS) stars. The stars are assumed to be close to the MS turn-off point in a globular cluster and are therefore modeled as $n = 3$, $\Gamma = 5/3$ polytropes. We find that the high degree of central mass concentration in these stars has a profound effect on the hydrodynamics. In particular, very little hydrodynamic mixing occurs between the dense, helium-rich inner cores and the outer envelopes. As a result, and in contrast to what has been assumed in previous studies, blue stragglers formed by direct stellar collisions are not necessarily expected to have anomalously high helium abundances in their envelopes or to have their cores replenished with fresh hydrogen fuel.

*Subject headings:* celestial mechanics, stellar dynamics – globular clusters: general – hydrodynamics – stars: blue stragglers – stars: interiors

[1] Center for Radiophysics and Space Research, Cornell University, Ithaca, NY 14853.

[2] Department of Astronomy, Cornell University.

[3] School of Natural Sciences, Institute for Advanced Study, Olden Lane, Princeton, NJ 08540. Email: rasio@guinness.ias.edu.

[4] Hubble Fellow.

[5] Department of Physics, Cornell University.

## 1. Introduction

It is now generally accepted that blue stragglers are formed by the merger of two main-sequence (hereafter MS) stars, either through direct physical collision or in a close binary system (Leonard 1989; Livio 1993; Stryker 1993; Bailyn & Pinsonneault 1995). Clear evidence for binary mergers has been found in the form of contact binaries among blue stragglers in the low-density globular clusters NGC 5466 (Mateo et al. 1990) and M71 (Yan & Mateo 1994), as well as in open clusters (Kaluźny & Rucinski 1993). Evidence for direct stellar collisions comes from recent detections by HST of large numbers of blue stragglers concentrated in the cores of very dense clusters such as M15 (De Marchi & Paresce 1994) and M30 (Yanny et al. 1994).

Benz & Hills (1987) performed the first three-dimensional calculations of parabolic collisions between two MS stars. An important conclusion of their study was that stellar collisions generally led to thorough mixing of the fluid. As a consequence, it was assumed in subsequent work that blue stragglers formed by stellar collisions should start their life close to the zero-age MS (hereafter ZAMS), but with an anomalously high He abundance in their envelopes. On the basis of this assumption, Bailyn (1992) suggested a way of distinguishing observationally between the two possible formation processes: blue stragglers made from collisions would exhibit higher He abundance than those made from binary mergers. Indeed, coalescence of two stars in a close binary is a much more gentle process than a direct collision (Rasio & Shapiro 1995), so that very little mixing is expected. More recently, Bailyn & Pinsonneault (1995) performed detailed stellar evolution calculations for blue stragglers. For the collisional case, they assumed chemically homogeneous initial models with enhanced He abundances, calculating the total He mass from the age of the cluster and the masses of the original MS stars.

In this *Letter*, we re-examine the question of mixing in stellar collisions. Benz & Hills (1987) used polytropic models with $n = 1.5$ and $\Gamma = 5/3$ to represent the MS stars. Unfortunately, such models apply only to very low-mass MS stars with large convective envelopes. For Population II MS stars, the effective polytropic index (defined in terms of the degree of central mass concentration) is close to $n = 1.5$ only for a mass $M \lesssim 0.4 M_\odot$ (see Lai, Rasio, & Shapiro 1994, Table 3). The object formed by the collision and merger of two stars with such low masses would hardly be recognizable as a blue straggler, since it would lie below, or not far above, the MS turn-off point in a CM diagram. Stars near the MS turn-off point have very shallow convective envelopes and are much better described by $n = 3$, $\Gamma = 5/3$ polytropes (Eddington's "standard model", see, e.g., Clayton 1983). Lai, Rasio & Shapiro (1993) have performed calculations of collisions between MS stars represented by this model, but they focused on the hyperbolic case relevant to galactic nuclei. Here we examine parabolic collisions, paying particular attention to the question of hydrodynamic mixing.

## 2. Numerical Method and Results

Our numerical calculations were done using the smoothed particle hydrodynamics (SPH) method (see Monaghan 1992 for a recent review). We used a modified version of the code developed by Rasio (1991) specifically for the study of stellar interactions (see Rasio & Shapiro 1991; 1992). SPH is a Lagrangian method, which should be ideally suited to the study of hydrodynamic mixing. Indeed, chemical abundances are passively advected quantities during dynamical evolution, so the chemical composition in the final fluid configuration can be determined very easily by simply noting the original positions of all SPH particles in the initial stellar models. For this study we have used the stellar evolution code developed by Sienkiewicz and collaborators (cf. Sienkiewick, Bahcall, & Paczyński 1990) to compute the chemical composition profile of the initial stellar models. We evolved a MS star of mass $M = 0.8\,M_\odot$, primordial He abundance $Y = 0.249$, and metallicity $Z = 0.001$ to the point of hydrogen exhaustion at the center and we used the interior profile of this terminal age MS (hereafter TAMS) model to assign values of the He abundance to all fluid particles. The age of this model is $t \simeq 15\,\text{Gyr}$, and the outer convective layer contains only about 0.2% of the mass.

The calculations reported here were done with $N = 3 \times 10^4$ equal-mass particles, each particle interacting with a constant number of neighbors $N_N \simeq 60$. We use time-dependent, individual particle smoothing lengths $h_i$ to ensure that the spatial resolution remains acceptable throughout the dynamical evolution. The gravity is calculated by a particle-mesh convolution algorithm, based on Fast Fourier Transforms (FFT) on a $128^3$ grid, which is readjusted at every timestep to follow the mass distribution in the system. A typical run takes about 100 CPU hours on an IBM SP-2 parallel supercomputer. Energy and angular momentum conservation is monitored throughout the integrations as a measure of numerical accuracy. In all of our calculations, this conservation was maintained to within a few percent.

Figure 1 illustrates the dynamical evolution for a typical case: two identical stars of mass $M$ were placed initially on a parabolic trajectory with pericenter separation $r_p = R$, where $R$ is the stellar radius. The stars begin with a separation of $5R$. The initial collision at time $t \simeq 5$ (in units of $(R^3/GM)^{1/2}$) disrupts the outer layers, but leaves the helium cores essentially undisturbed. The stars withdraw to apocenter at $t \simeq 27$, and by $t \simeq 50$ are colliding for the second time. There are a total of four such interactions before the stars merge completely. The final equilibrium configuration (Fig. 2) is an axisymmetric, rapidly rotating object. The density profile is steeper than that of an $n = 3$ polytrope. About 1% of the total mass has been lost after the collision. The results will be discussed in more detail in Lombardi, Rasio, & Shapiro (1995a), where we will also present calculations done for other values of $r_p/R$ and different combinations of masses.

Figure 3 shows the He mass fraction $Y$ as a function of the interior mass fraction $m/M$ (where $m$ is the mass inside a isodensity surface, and $M$ is the total mass of the bound fluid) for the final merged configuration. The points correspond to the final ($t = 85$) particle values, with the long-dashed curve representing their average. Also shown for comparison are the profiles corresponding to a completely mixed merger remnant (horizontal line), to a TAMS star with the same total mass (short-dashed curve), and to the initial TAMS stars (solid curve). It is immediately apparent that the He enrichment in the outer layers is minimal. The outer 20% of the mass has an average value of $Y = 0.250$, just barely above the assumed primodial value of $Y = 0.249$. Hydrogen enrichment in the core is also minimal, with the innermost 20% of the mass being 82% helium. In addition, the object has $Y \simeq 1$ at the center and contains a total amount of He ($M_{\rm He} = 0.66\,M_\odot$) which is larger than that of a normal TAMS star with the same total mass, primordial He abundance, and metallicity ($M_{\rm He} = 0.57\,M_\odot$).

## 3. Discussion

The main reason for the negligible hydrodynamic mixing is that the He is concentrated in the high-density inner cores of the initial stars (90% of the He produced during the MS lifetime of a $0.8\,M_\odot$ star is contained in the inner 1% of the volume, and the mean density inside this region is 42 times higher than outside). These high-density cores are very difficult to disrupt, even in a direct collision. In the final merger, the helium-rich material, which is also of lower specific entropy, tends to remain concentrated near the bottom of the gravitational potential well.

It must be stressed that the amount of mixing determined by SPH calculations is actually an *upper limit*. Indeed, some of the mixing observed in a calculation will always be a numerical artifact. Low-resolution SPH simulations in particular tend to be very noisy (cf. Rasio & Shapiro 1991), and the noise can lead to spurious diffusion of SPH particles, independent of any real physical mixing of fluid elements. This may be what caused the apparently thorough mixing of SPH particles observed by Benz & Hills (1987) for grazing collisions (which tend to require very long integrations). For more nearly head-on collisions, our results are in qualitative agreement with those of Benz & Hills (1987) in predicting a negligible amount of hydrodynamic mixing.

We have performed a series of systematic tests to evaluate quantitatively the effects of spurious transport in SPH calculations (Lombardi et al. 1995b). The results suggest that much of the mixing observed in our calculations could in fact have been caused by spurious diffusion of SPH particles. One such test is particularly relevant. We can exploit the reflection symmetry of the problem: in the absence of spurious diffusion, SPH particles close to the orbital plane at $t=0$ should remain close to this plane throughout the evolution. By measuring their actual displacement away from the orbital plane, we can estimate the effects of spurious diffusion. For a calculation using the parameters of §2 we find that these displacements corresponds to a root mean square change in the interior mass fraction of $(\Delta m/M)_{rms} \simeq 0.1$ over the entire simulation. This is very close to the root mean square change measured for all particles in the simulation of §2, $(\Delta m/M)_{rms} \simeq 0.11$. Thus the real hydrodynamic mixing of fluid elements in a direct collision may be even smaller than what is shown in Figure 3.

The results described here are typical of all encounters with $0 \leq r_p/R \lesssim 1$ (Lombardi et al. 1995a). More grazing encounters (with $r_p/R \gtrsim 1$) of $n=3$ polytropes cannot be computed directly with SPH because the orbital energy dissipated during the first close interaction is very small ($\Delta E \simeq 0.02\,GM^2/R$ for $r_p/R = 1.5$) and the integration time until the next pericenter passage can be several orders of magnitude larger than the hydrodynamic time. These encounters lead to the formation of a binary system, and are better described as "tidal captures" than "collisions." The ultimate fate of tidal-capture binaries is a rather controversial question (Kochanek 1992; Mardling 1995), but it seems very likely that at least all encounters with $r_p/R \lesssim 2$ will ultimately lead to mergers. Extrapolation from the results obtained for $0 \leq r_p/R \lesssim 1$, taking into account the importance of spurious diffusion, indicates that hydrodynamic mixing remains always negligible in mergers (Lombardi et al. 1995a). However, we cannot rule out the possibility that other processes taking place on a thermal timescale (e.g., meridional circulation or convection) could still lead to significant mixing as the merged object recontracts to thermal equilibrium.

We thank R. Sienkiewicz for providing us with a copy of his stellar evolution code. Support for this work was provided by NSF Grant AST 91-19475 and NASA Grant NAG 5-2809 to Cornell University. F. A. R. is supported by a Hubble Fellowship, funded by NASA through Grant HF-1037.01-92A from the Space Telescope Science Institute, which is operated by AURA, Inc., under contract NAS5-26555. Computations were performed at the Cornell Theory Center, which receives major funding from the NSF and IBM Corporation, with additional support from the New York State Science and Technology Foundation and members of the Corporate Research Institute.

Fig. 1.— Snapshots of particle positions at various times during the parabolic collision of two MS stars (modeled as $n = 3$, $\Gamma = 5/3$ polytropes) initially on an orbit with periastron separation $r_p = R$. All particles have been projected onto the orbital plane. Time is measured in units of $(R^3/GM)^{1/2}$ and distance in units of $R$. See text for details.

Fig. 2.— Pressure contours and velocity field in the (a) orbital plane and (b) y=0 plane for the final ($t = 85$) configuration of the collision shown in Fig. 1. There are 15 pressure contours, which are spaced logarithmically and cover 5 decades down from the maximum.

Fig. 3.— Helium mass fraction $Y$ as a function of interior mass fraction. The points correspond to the final particle values and the long-dashed curve represents their average. Also shown for comparison are the profiles corresponding to a completely mixed merger remnant (horizontal line), a TAMS star with the same total mass (short-dashed curve), and the initial TAMS stars (solid curve).

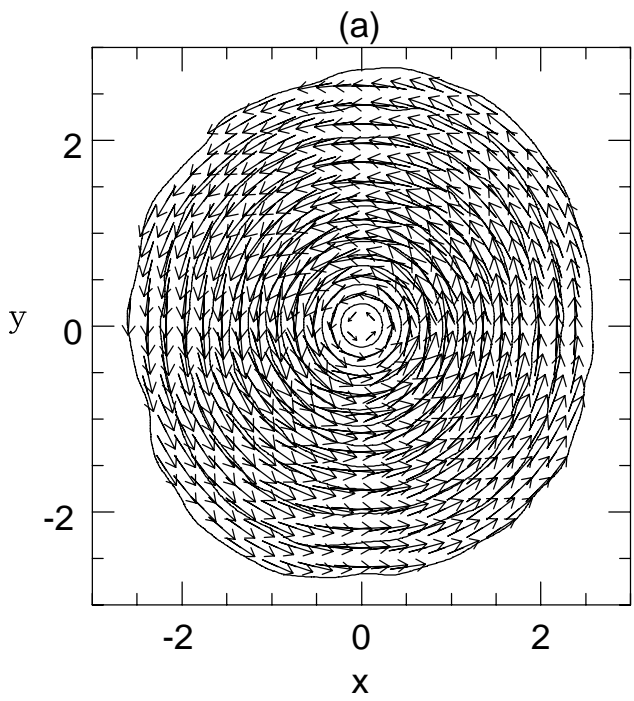 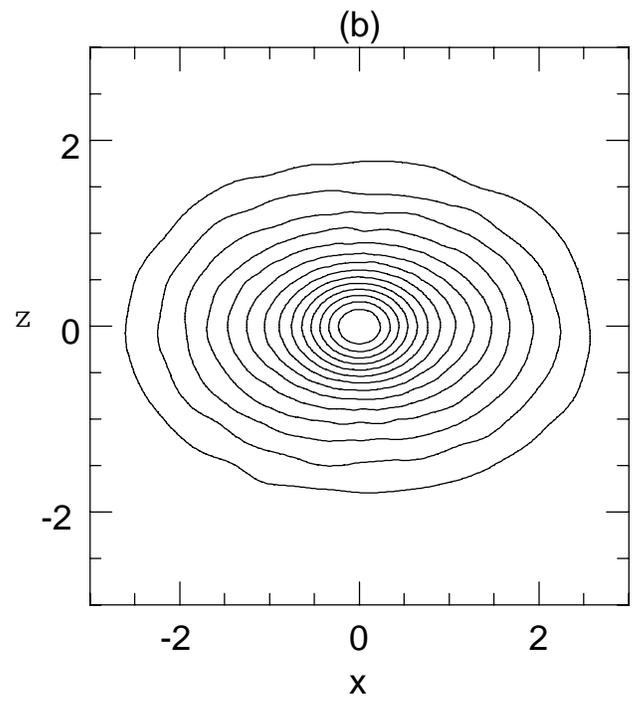

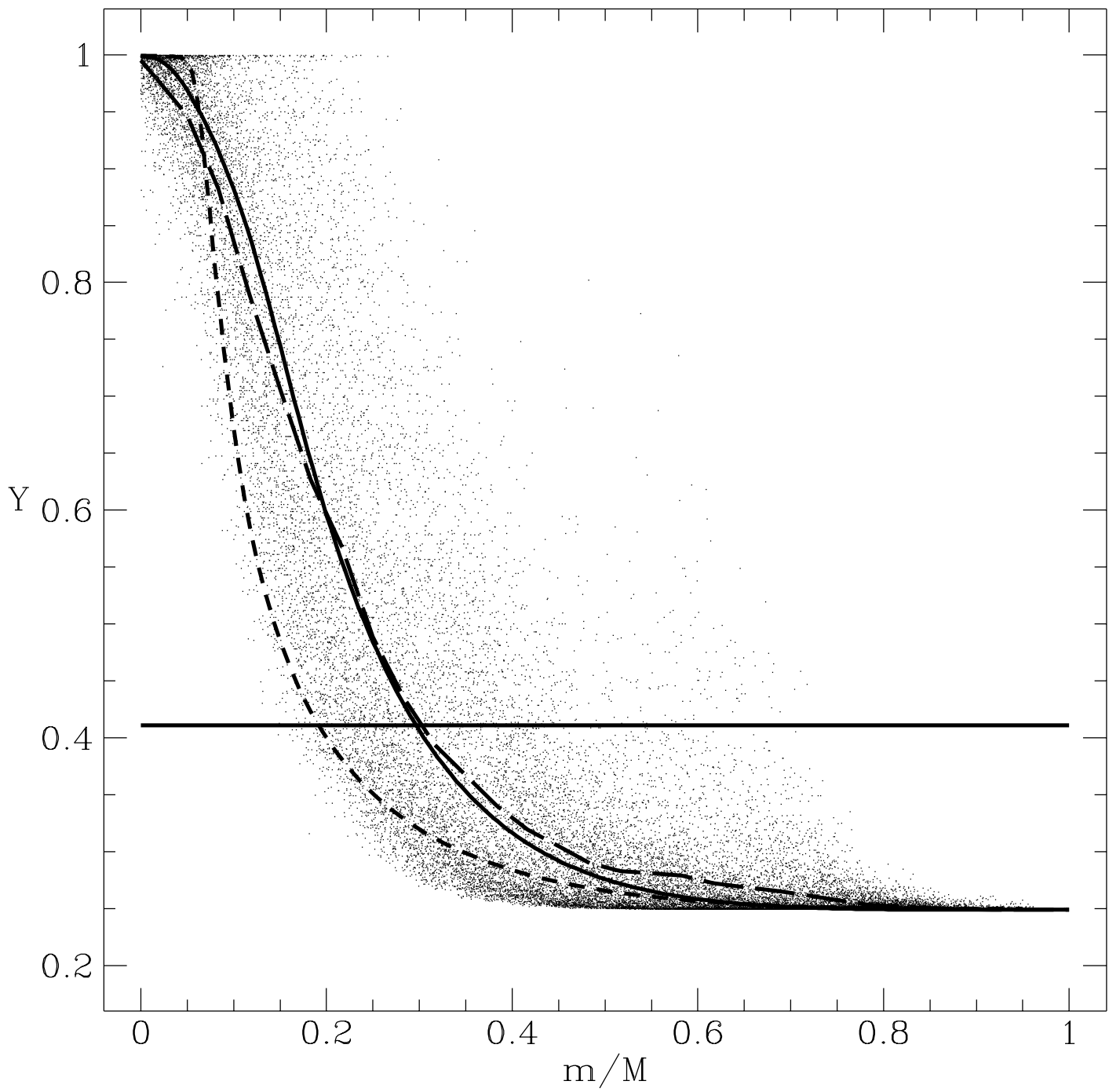